# A Positioning System in an Urban Vertical Heterogeneous Network (VHetNet)

Hongzhao Zheng, *Member, IEEE*, Mohamed Atia, *Senior Member, IEEE,* and Halim Yanikomeroglu, *Fellow, IEEE*

*Abstract*—Global navigation satellite systems (GNSSs) are essential in providing localization and navigation services to most of the world due to their superior coverage. However, due to high pathloss and inevitable atmospheric effect, the positioning performance of any standalone GNSS is still poor in urban areas. To enhance the positioning performance of legacy GNSSs in urban areas, a positioning system, which utilizes high altitude platform station (HAPS) and 5G gNodeBs (gNBs), in a futuristic urban vertical heterogeneous network (VHetNet) is proposed. In this paper, we demonstrate the effectiveness of gNBs in improving the vertical positioning performance for both the GPS-only system and the HAPS-aided GPS system by analyzing the impact of the density of gNBs and the pseudorange error of gNB on the positioning performance of the gNB augmented positioning systems. We also demonstrate the effectiveness of receiver autonomous integrity monitoring (RAIM) algorithms on the HAPS and/or gNB aided GPS systems in urban areas.

## I. INTRODUCTION

Global navigation satellite systems (GNSS) have been widely used for localization and navigation since 1993 when the 24-satellite system of the global positioning system (GPS) became fully operational. Although relevant research has been continuously carried out since then, any standalone GNSSs can only offer a meter level global average civilian positioning accuracy to date [1], not to mention the performance in urban areas where signal qualities can be degraded severely due to blockages and the multipath effect. Due to the fact that Earth is not transparent, and we can only receive signals above the horizon, the diversity in the elevation angle would not be as rich as that in the azimuth angle. In consequence, the vertical positioning performance of legacy GNSSs tends to be worse than the horizontal positioning performance. Even though vertical positioning accuracy is less crucial than horizontal positioning accuracy today, it might become much more critical in the near future. For example, as urbanization is continuously developing, we can expect more multi-layered overpass highways to be constructed in urban areas. Therefore, it is imperative to improve the vertical positioning accuracy so that vehicles can be localized and navigated accurately when driving on multi-layered overpass highways. Another potential application is the unmanned aerial vehicles (UAVs) maneuvering in aerial highways [2], which would require precise positioning services for navigation, guidance, and control purposes.

There are many challenges for the localization in urban areas, such as ranging source availability, signal quality, and so forth. One way to improve the outdoor positioning performance in urban areas is to use a multi-constellation GNSS, which provides more visible satellites for positioning. For instance, using a city model of London with decimeter-level accuracy and the dilution of precision (DOP) as the metric for the positioning performance evaluation, Wang et al. show that the positioning performance of the multi-constellation GNSS, which consists of GPS, GLONASS, Galileo, and Compass, is much better than only using GPS and GLONASS [3]. With integrated GLONASS and GPS data instead of GPS data alone collected under a strong ionospheric scintillation condition, Marques et al. demonstrate that up to 60% positioning accuracy improvement can be achieved using the precise point positioning (PPP) method [4].

There are thousands of operational low-Earth-orbit (LEO) satellites in space today, providing broadband communication services, such as global maritime coverage and integrated satellite-aerial communications [5]. It is reported that there are 3236 operational Starlink LEO satellites in orbit as of November 2022. Furthermore, it is expected that there will be tens of thousands of LEO satellites launched into space by 2030 [5]. Compared to the typical type of satellite deployed in GNSSs, which is the medium-Earth-orbit (MEO) satellites, LEO satellites bring advantages such as less pathloss and lower latency owing to shorter distance to Earth, and greater coverage owing to the abundant quantity. Therefore, utilizing LEO satellite signals has been considered and proved to be another viable way to improve the positioning performance of legacy GNSSs. For example, Meng et al. show that the convergence time of PPP can be significantly reduced from 30 minutes by using a standalone GNSS to shorter than 5 minutes by using a LEO global navigation augmentation system [6]. By

This paper was supported in part by Huawei Canada.
H. Zheng is with the Embedded and Multi-sensor Systems Lab (EMSLab) and the Non-Terrestrial Networks (NTN) Lab, Department of Systems and Computer Engineering, Carleton University, Ottawa, ON, K1S 5B6, Canada (e-mail: hongzhaozheng@cmail.carleton.ca).
M. Atia is with the Embedded and Multi-sensor Systems Lab (EMSLab), Department of Systems and Computer Engineering, Carleton University, Ottawa, ON, K1S 5B6, Canada (e-mail: Mohamed.atia@carleton.ca).
H. Yanikomeroglu is with the Non-Terrestrial Networks (NTN) Lab, Department of Systems and Computer Engineering, Carleton University, Ottawa, ON, K1S 5B6, Canada (e-mail: halim@sce.carleton.ca).

introducing 60, 96, 192, and 288 polar-orbiting LEO constellations to the nominal GPS constellation, Li et al. show that the real-time kinematic (RTK) convergence time for a 68.7 km baseline can be reduced from 4.94 to 2.73, 1.47, 0.92, and 0.73 min, respectively; and its corresponding average time to first fix (TTFF) can be shortened from 7.28 to 3.33, 2.38, 1.22, and 0.87 min, respectively [7].

Navigation and localization using satellites have been successful in the sense that satellites can provide wide coverage to the users on the ground. With a number of them, almost all the inhabited areas can be well covered by satellites. However, there are a few disadvantages of using satellites. For instance, the utilization efficiency of satellites is low since about 70.8% of the surface of Earth is covered by the ocean [8]. Due to the fast-moving nature of satellites, LEO satellites in particular, users will experience frequent handovers which require proper management. In terms of GPS time synchronization, although the atomic clocks on satellites are highly accurate, we still need to account for the time dilation incurred due to the fast-moving nature of satellite, especially LEO satellites, according to the special theory of relativity. Apart from this, we should also consider the time dilation incurred due to the weak gravity on a distant satellite according to the general theory of relativity [9].

To address the aforementioned issues, high altitude platform station (HAPS [1]), which, according to the international telecommunication union (ITU), is defined as "radio stations located on an object at an altitude of 20-50 kilometers and at a specified, nominal, fixed point relative to the Earth" [10], can be considered. By this definition, there are many use cases for HAPS, such as the backhauling of small and isolated base stations (BSs), using HAPS as aerial data centers, and LEO satellite handoffs management [11]. As HAPS are closer to the Earth surface, the pathloss for HAPS signals will be less than that for satellite signals. Considering a similar transmitted power on both HAPS and satellite signals, we can expect the received signal power for HAPS signals to be greater than that for satellite signals, as a result of which the bit error rate (BER) for HAPS signals is expected to be less than that for satellite signals. As the residual errors in the pseudorange equation can be subject to the precision of the correlation between the transmitted navigation code and the locally generated code in the receiver, which in turn depends on the strength of the transmitted signal, a stronger signal will likely render a clearer navigation code, leading to a more precise range estimation. Therefore, the parameter estimation errors of the pseudorange equation for the HAPS are likely less than that for satellites. In addition, the time dilation of the atomic clocks on the HAPS should be negligible in accordance with the special theory of relativity, and significantly diminished in accordance with the general theory of relativity. Consequently, the atomic clocks on HAPS will maintain a higher precision compared to those on satellites. Therefore, the HAPS clock offset will likely be estimated with an error less than that for the satellite clock offset. To minimize the effects of atmospheric turbulence and weather conditions and comply with the airspace regulations for the maximum altitude for UAVs in many countries, the typical altitude considered for HAPS is around 20 km. HAPS at this altitude are well below the ionosphere, therefore their signals can be expected to have no ionospheric delay. We should note that the ionospheric delay is known as one of the largest errors in pseudorange measurements [12]. Positioning error of the ranging source contributes to the overall estimation error, thereby degrading positioning performance. Like satellite orbital error, HAPS positioning error should be considered in receiver position computation. There are a few papers focusing on the positioning of HAPS in literature. For example, a 0.5 m positioning accuracy (circular error probable [CEP] 68%) for a HAPS is shown to be achievable using the modified RTF method [13]. With the advantages offered by HAPS, we can deploy HAPS, which functions as ranging sources with positioning payload such as satellite-grade atomic clocks, on top of metro cities where user demand is likely the highest.

5G new radio (NR) positioning which involves the use of the 5G wireless communication infrastructure has been a popular topic in both academia and industry in recent years. With 3GPP Release 16 specifications, many enterprises have attempted to either simulate realistically or trial physically the 5G NR positioning. For example, Dwivedi et al. from Ericsson incorporate the 3GPP simulation scenarios and have shown that a 90-percentile outdoor positioning accuracy of a few meters down to a few decimeters is achievable depending on the simulation scenario [14]. The 5G NR single-cell positioning prototyped jointly by Qualcomm and ZTE has showcased an 80-percentile horizontal accuracy of 2.1 m and an 80-percentile vertical accuracy of 1.2 m based on round trip time (RTT) and angle of arrival (AOA) [15]. These results not only prove that 5G infrastructures and technologies, such as gNodeB (gNB), high frequency band, and massive multi-input multiple-output (MIMO) antenna, are capable of providing positioning services with performance comparable to or even better than satellites, but also show that gNBs have the potential to improve vertical positioning accuracy. The prominent positioning performance of using gNBs can be explained by the fact that users are served and tracked by gNBs with beam steering. It not only improves the signal-to-noise-ratio (SNR), but also helps mitigate the multipath effect and improve BER by directing the strongest signal in the direction of the receiver. Therefore, with gNBs augmenting GNSSs at relatively low elevation angles, we would expect a greater positioning availability with an improved vertical dilution of precision (VDOP), or in other words, vertical positioning accuracy.

To take advantage of existing vertical heterogeneous networks and a futuristic aerial network in the stratosphere, we propose a novel positioning system in an urban vertical heterogeneous network (VHetNet) that involves satellites, HAPS, and 5G gNBs. Although an increase in complexity might occur due to the integration of aerial, space, and terrestrial signals, there are numerous approaches to solve the complexity increase issue due to data fusion. One of the techniques for complexity reduction is known as dimensionality reduction, such as principal component analysis (PCA) and physical features extraction, which can effectively reduce the dimensionality of the data with linear relationships at the cost of a minor loss of information [16]. Another approach for complexity reduction could involve the utilization of receiver autonomous integrity monitoring (RAIM) algorithms which are known to be effective in detecting and excluding signals that are deemed to be of low quality. Based on the characteristics of HAPS, we assume that the overall parameter estimation error of the pseudorange equation for a HAPS is less than that for a satellite. Hereafter for convenience, the overall parameter estimation error of the pseudorange equation is rephrased as the pseudorange error. Owing to the short distance between a gNB and a receiver and the fact that the position of a gNB can be precisely known, we could expect a higher received power with negligible tropospheric error on a gNB signal in comparison with a HAPS signal. Therefore, the pseudorange error of gNB will likely be smaller than that of HAPS. This assumption is also supported by the positioning performance of using gNBs publicized by the industry [14][15].

---

[1] In this paper, the word "HAPS" is used for both singular and plural forms. This is in accordance with the convention adopted in the International Telecommunications Union (ITU) documents.

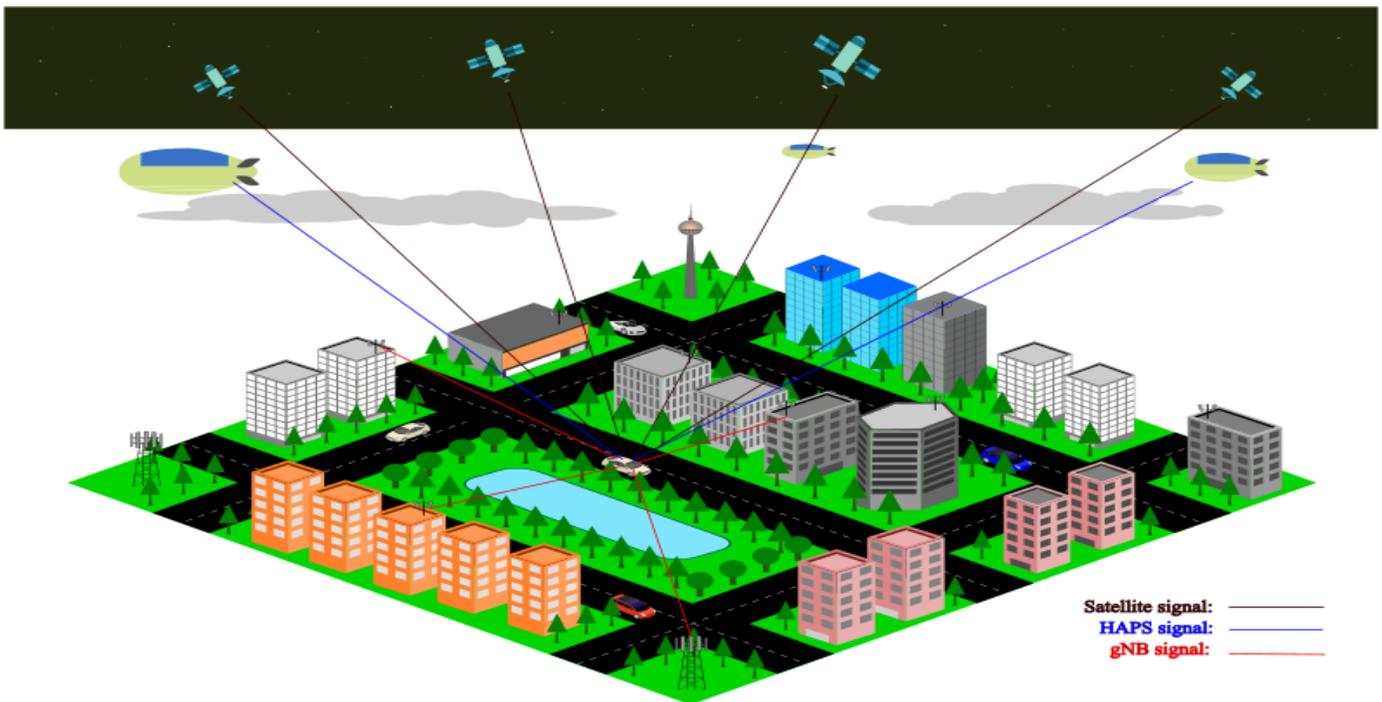

Fig. 1. System model of a positioning system in an urban VHetNet.

In urban areas, signals tend to come with a reduced availability and a degraded quality, because they are highly subject to blockages and multipath effect. Applying a signal selection algorithm, which can exclude signals of poor quality, can further enhance the positioning performance of the augmented GNSS. One of the commonly used signal selection algorithms is the RAIM algorithm. For example, using a RAIM algorithm along with reliability and separability methods, Hewitson and Wang show that a deviation of 20 m can be detected with 80% probability at the 0.5% significance level for the GPS/GLONASS/Galileo constellation [17]. By carrying out both a static and a pedestrian test in urban areas using a GPS/GLONASS standard receiver, Angrisano et al. show that the use of RAIM algorithms can effectively reduce the root mean square (RMS) and maximum errors for both the vertical and horizontal component, demonstrating the effectiveness of the use of RAIM for the augmented GNSS in urban areas [18]. As the proposed positioning system considers using gNBs and HAPS to augment a legacy GNSS in urban areas, we also would like to examine the effectiveness of RAIM algorithms on the HAPS and/or gNB aided GPS systems.

There are four main positioning systems considered in this work, namely the GPS-only system, HAPS + GPS system, gNB + GPS system, and gNB + HAPS + GPS system. For the interest of a research problem, the pseudorange error of gNB is simulated using a range of values, all of which are smaller than that for HAPS. To examine the impact of the density of gNBs on the positioning performance of the gNB augmented positioning systems, we simulate three scenarios with varied number of gNBs. Since both HAPS and gNB measurements are unavailable to date, they are simulated using reasonable assumptions and accredited documentations. The HAPS positions used in this work are carefully simulated using a professional GNSS simulator. We would like to acknowledge the following as the same in our prior work [19]: 1) the choice of the pseudorange errors of HAPS in the suburban and urban scenarios; 2) real GPS data; 3) simulated HAPS positions; 4) the single point positioning (SPP) algorithm. The contributions of this paper are listed below.

- In this work, we propose a novel positioning system that

- utilizes several components in an urban VHetNet, such as GPS satellites, HAPS, and 5G gNBs. First, we describe the modeling of the pseudorange for gNBs using reasonable assumptions; and we implement the line-of-sight (LOS) probability model for the gNBs in urban areas on the basis of a 3GPP technical report (TR).
- Secondly, with real GPS data collected by GNSS receivers, and the HAPS and gNB data simulated using a professional GNSS simulator, we analyze the impact of the density of gNBs on both the horizontal and vertical positioning accuracies of the gNB augmented positioning systems, from which we prove the effectiveness of gNBs in improving the vertical positioning accuracy and HAPS in improving both the horizontal and vertical positioning accuracy.
- Thirdly, we analyze the impact of the pseudorange error of gNB on both the horizontal and vertical positioning accuracies of the gNB augmented positioning systems, from which we again prove the effectiveness of gNBs in improving the vertical positioning accuracy.
- Finally, we demonstrate the effectiveness of RAIM algorithms on the HAPS and/or gNB aided GPS systems in urban areas.

The rest of the paper is organized as follows. Section II describes the system model, the simulation setup, and the line-of-sight (LOS) probability model for both HAPS and gNB. Section III provides the impact of the density of gNBs and the pseudorange error of gNB on the positioning performance of the gNB augmented positioning systems, and the effectiveness of RAIM on the HAPS and/or gNB aided GPS systems in urban areas, consecutively. Finally, Section IV presents conclusions and future research directions.

## II. System Model

The system model of the proposed positioning system in an urban VHetNet is depicted in Fig. 1. As we can see, a vehicle (receiver) is moving in an urban area and is receiving signals from entities spanning from terrestrial network, aerial network, and space network. As the region of interest is urban areas, being able to receive signals from at least four satellites, which is the minimum number required to conduct a 3D localization, with a standalone GNSS is not guaranteed. Direct LOS links between ranging sources and the user might not be established due to blockages, such as skyscrapers and vegetations. Moreover, some of the received signals might have much lower power due to the multipath effect. To complement a legacy GNSS for localization, HAPS and gNB signals are simultaneously utilized. By so doing, it not only increases the ranging source availability, but also provides a few reliable ranging signals for localization. As the atmospheric error due to signal refraction is negligible for a satellite whose elevation is above 15 degrees [20], which can likely be generalized to a HAPS. Therefore, we decided to use an elevation mask of 15 degrees for both satellites and HAPS. Note that we do not discard the gNB signals that are received from very low

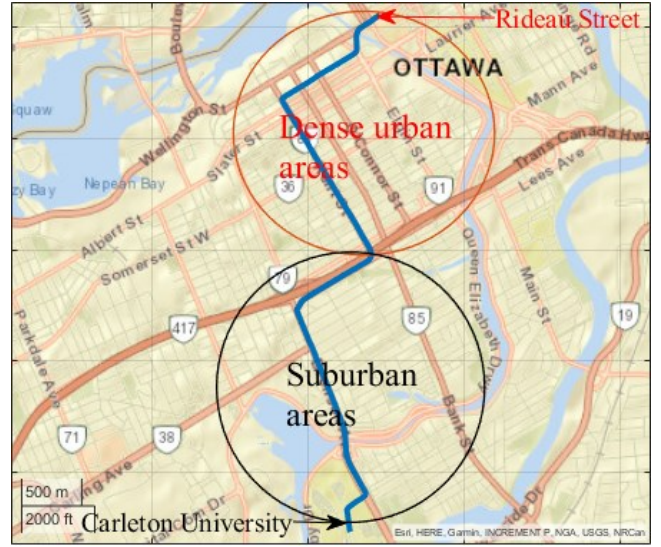

Fig. 2. Vehicle trajectory when collecting real GPS data.

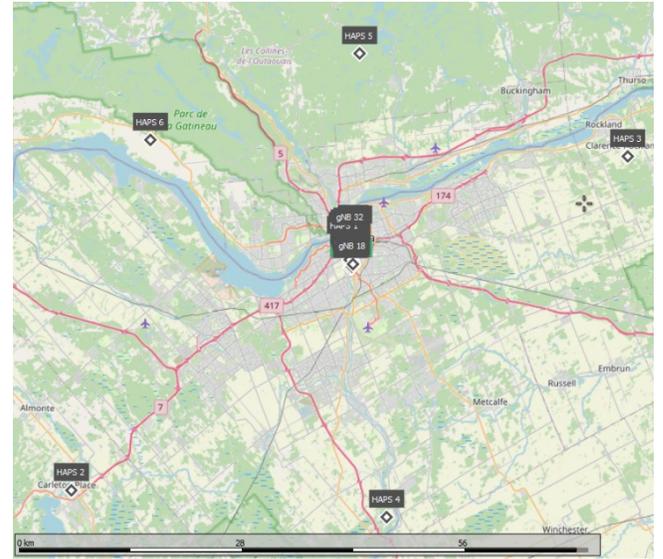

Fig. 3. Simulated HAPS locations.

elevation angles (e.g., <10 degrees), because the atmospheric error on them is negligible.

The pseudorange equation for satellite is as follows:

$$p_{SAT} = \rho_{SAT} + d_{SAT} + c(dt - dT_{SAT}) + d_{ion,SAT} + d_{trop,SAT} + \epsilon_{mp,SAT} + \epsilon_p \quad (1)$$

where $p_{SAT}$ is the pseudorange measurement for satellite, $\rho_{SAT}$ represents the geometric range between satellite and receiver, $d_{SAT}$ denotes the satellite orbital error, $c$ denotes the speed of light, $dt$ denotes the receiver clock offset, $dT_{SAT}$ denotes the satellite clock offset, $d_{ion,SAT}$ represents the ionospheric delay for satellite signals, $d_{trop,SAT}$ represents the tropospheric delay for satellite signals, $\epsilon_{mp,SAT}$ represents the multipath delay for satellite signals, and $\epsilon_p$ denotes the receiver noise induced delay. The pseudorange equation for HAPS is expressed by

$$p_{HAPS} = \rho_{HAPS} + d_{HAPS} + c(dt - dT_{HAPS}) + d_{trop,HAPS} + \epsilon_{mp,HAPS} + \epsilon_p \quad (2)$$

where $p_{HAPS}$ is the pseudorange measurement for HAPS, $\rho_{HAPS}$ denotes the geometric range between HAPS and the receiver, $d_{HAPS}$ denotes the HAPS position error, $dT_{HAPS}$ denotes the HAPS clock offset, $d_{trop,HAPS}$ represents the tropospheric delay for HAPS signals, and $\epsilon_{mp,HAPS}$ denotes the multipath delay for HAPS signals. We would like to acknowledge that the pseudorange equations for satellites and HAPS are the same as in [19]. Based on the characteristics of the gNB signal, the pseudorange equation for gNB is expressed as follows:

$$p_{gNB} = \rho_{gNB} + c(dt - dT_{gNB}) + \epsilon_{mp,gNB} + \epsilon_p \quad (3)$$

where $p_{gNB}$ is the pseudorange measurement for gNB, $\rho_{gNB}$ represents the geometric range between gNB and the receiver, $dT_{gNB}$ denotes the gNB clock offset, and $\epsilon_{mp,gNB}$ denotes the multipath delay for gNB signals. In this paper, the horizontal dilution of precision (HDOP), the VDOP, the horizontal positioning accuracy, and the vertical positioning accuracy are used as the key performance indicators (KPIs) for the analyses of interest. To calculate the HDOP, the covariance matrix in the local north-east-down (NED) reference frame should be found as follows:

$$\boldsymbol{Q}_{NED} = \begin{bmatrix} \sigma_n^2 & \sigma_{ne} & \sigma_{nd} \\ \sigma_{ne} & \sigma_e^2 & \sigma_{ed} \\ \sigma_{nd} & \sigma_{ed} & \sigma_d^2 \end{bmatrix}. \quad (4)$$

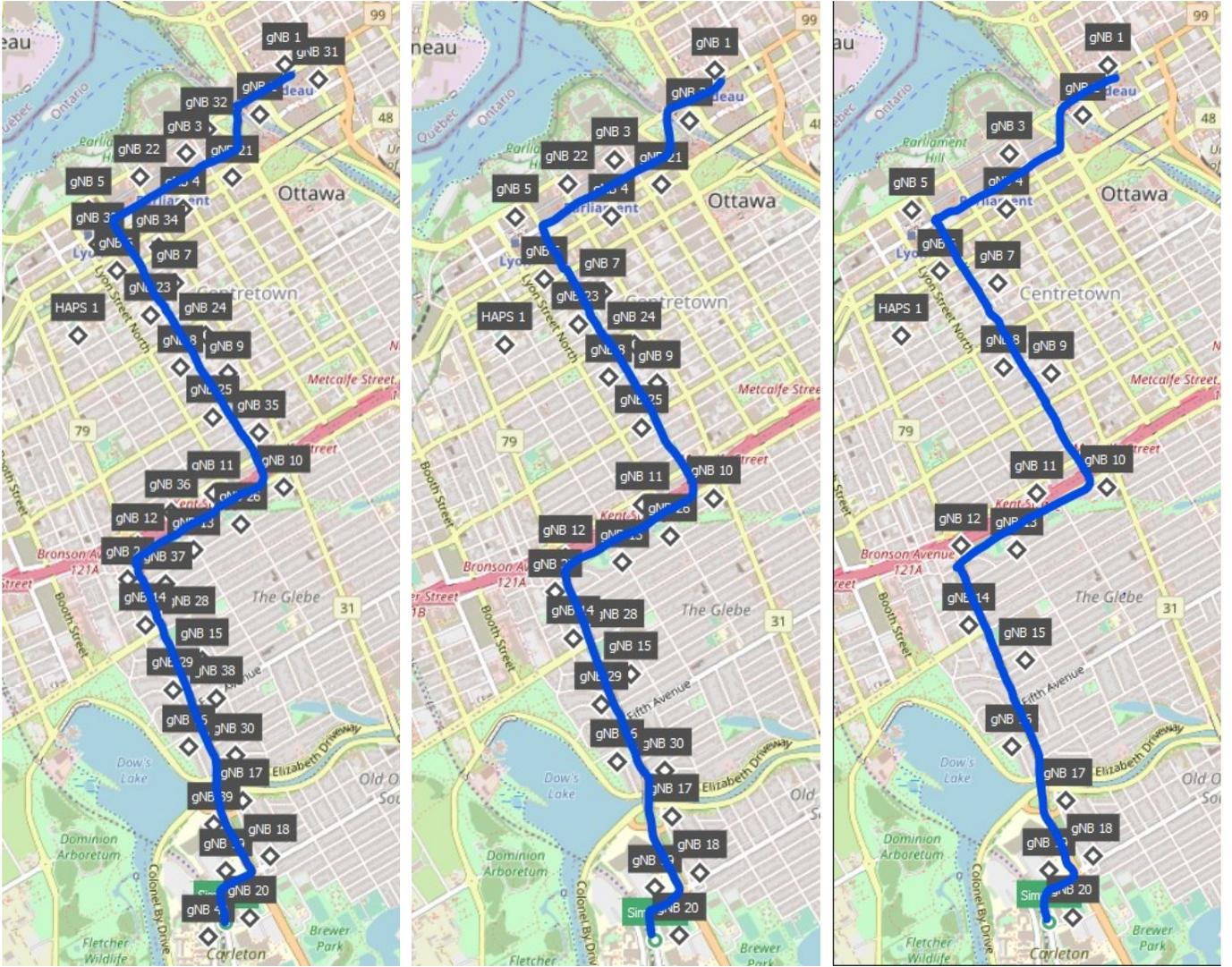

Fig. 4. Locations of the simulated gNBs (40-gNB: left; 30-gNB: middle; 20-gNB: right; the vehicle trajectory is in blue).

Then, the HDOP is described by

$$HDOP = \sqrt{\sigma_n^2 + \sigma_e^2} \quad (5)$$

and the VDOP is described by

$$VDOP = \sqrt{\sigma_d^2} \quad (6)$$

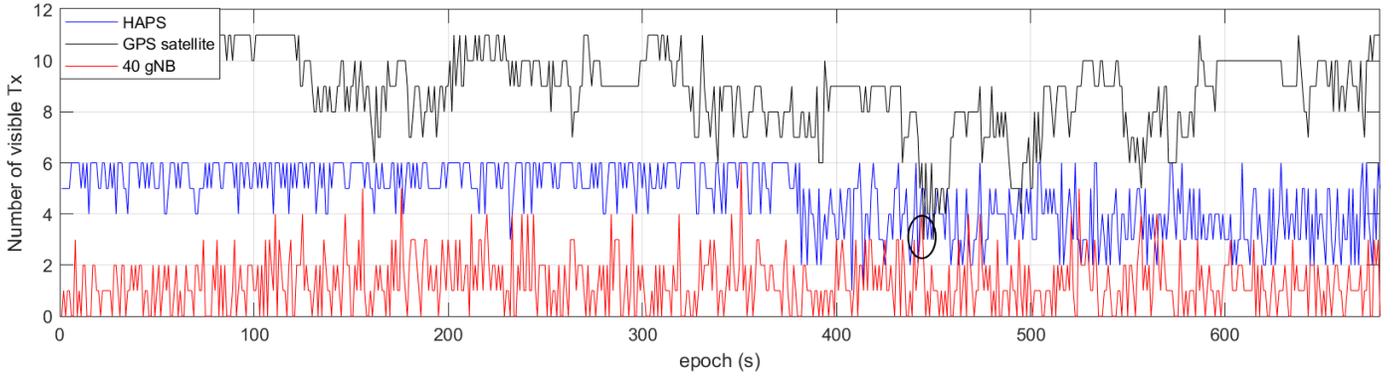

Fig. 6. Availability of GPS satellite, HAPS, and gNB during the observation period.

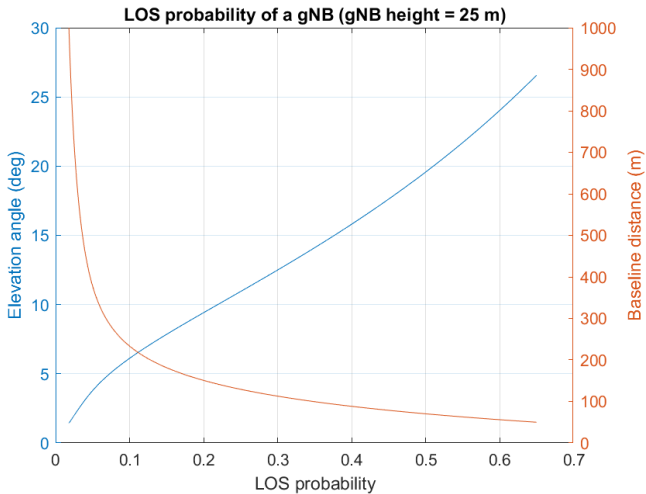

Fig. 5. LOS probability of a gNB with respect to elevation angle and baseline distance.

where $\sigma_n$, $\sigma_e$, and $\sigma_d$ denote the error on the receiver position in the local north, east, and down directions, respectively. Since HAPS is a futuristic object and gNB has not been fully deployed to date, their positions are carefully simulated using the Skydel GNSS simulator [21]. The pseudorange of both HAPS and gNB is modeled as follows:

$$p = \rho + \sigma_p \quad (7)$$

where $\sigma_p$ denotes the pseudorange error which follows the Gaussian distribution with zero mean and varied standard deviations. We follow the same way as in our prior work [19] in choosing the pseudorange error of HAPS: 2 m in suburban area and 5 m in urban area. The pseuodrange error for gNB is simulated with three values: 0 m, 0.5 m, and 1.5 m, all of which are smaller than the pseudorange error of HAPS. Due to the stationarity of HAPSs and their short distances to the ground of Earth, the time dilation of the atomic clocks on HAPS is negligible, thereby zero HAPS clock offset is considered in this work. The gNB clock offset is also set to zero for simplicity. We use the SPP algorithm to compute the receiver position. A RAIM algorithm based on [22] is implemented to further improve the positioning performance of the considered positioning systems. An optimized signal design, such as high chipping rate, can be effective in improving the positioning performance from several aspects. For example, GPS L5 has been shown to be effective in reducing the TTFF ambiguity [23] and mitigating the multipath effect [24]. As this work is concerned with a system-level simulation, only GPS C/A L1 signal is considered for simplicity.

### A. Simulation Setup

The vehicle trajectory that we followed to collect real GPS data is shown in Fig. 2. As we can see, we depart from Carleton University in a suburban area and arrive at Rideau Street in an urban area of Ottawa. We simulate six HAPS whose locations are depicted in Fig. 3. From this figure, we see that one HAPS is placed above downtown Ottawa and the others are positioned over nearby populated aeras and conservation areas. All the simulated HAPS are positioned 20 km about the ground and follow a circular trajectory with a radius of 300 m. HAPS are positioned in this way such that they constitute a near Zenith + Horizon (ZH) geometry, providing a satisfying DOP [25]. To examine the impact of the density of gNBs on the positioning performance of the gNB-aided positioning systems, namely the gNB + GPS system and the gNB + HAPS + GPS system, we simulate three scenarios where each scenario considers a different number of gNBs: 20 gNBs, 30 gNBs, and 40 gNBs. We use the coverage of the 5G mmWave type gNBs provided in [26], which considers the network dimensioning requirements for 5G fixed wireless access (FWA) implementation, in choosing the locations for gNBs in the 40-gNB scenario. All the simulated gNBs have the same height of 25 m. The locations of gNBs are shown in Fig. 4. As we can see that the locations of gNBs are distributed along the vehicle trajectory. With the topology of the simulated gNBs, the closest distance from any gNB to the vehicle trajectory is found to be around 50 m.

### B. Line-of-sight (LOS) Probability Models for HAPS and gNB

To implement a realistic LOS probability for HAPS in our simulation, we follow the LOS probability model for the HAPS proposed in [27] and [28]. We follow the LOS probability model for the UMa scenario, which is documented in 3GPP TR 38.901 Release 16 [29], in implementing the LOS probability for gNBs in urban areas. Since the height of the receiver in this work is about 2 m, therefore the LOS probability model for the UMa scenario is the same as that for the UMi-Street Canyon scenario. The LOS probability model for gNBs is as follows:

$$Pr_{LOS} = \begin{cases} 1 & , d_{2D} \leq 18m \\ \frac{18}{d_{2D}} + \exp\left(-\frac{d_{2D}}{36}\right)\left(1 - \frac{18}{d_{2D}}\right) & , 18m < d_{2D} \end{cases} \quad (7)$$

where $Pr_{LOS}$ denotes the LOS probability for gNB, and $d_{2D}$ represents the baseline distance between a gNB and the receiver. Based on the LOS probability model, a plot of the LOS probability as a function of the elevation angle and the baseline distance to a gNB is shown in Fig. 5. As we can see, under the simulation setup described in the previous subsection, all gNB signals are received with a baseline distance longer than 50 m and an elevation angle lower than 26.57°. Notice that the availability of gNBs can be increased if we raise the height of gNBs. However, signals received from a higher elevation angle would have a reduced improvement on the VDOP/vertical positioning accuracy and a higher pathloss.

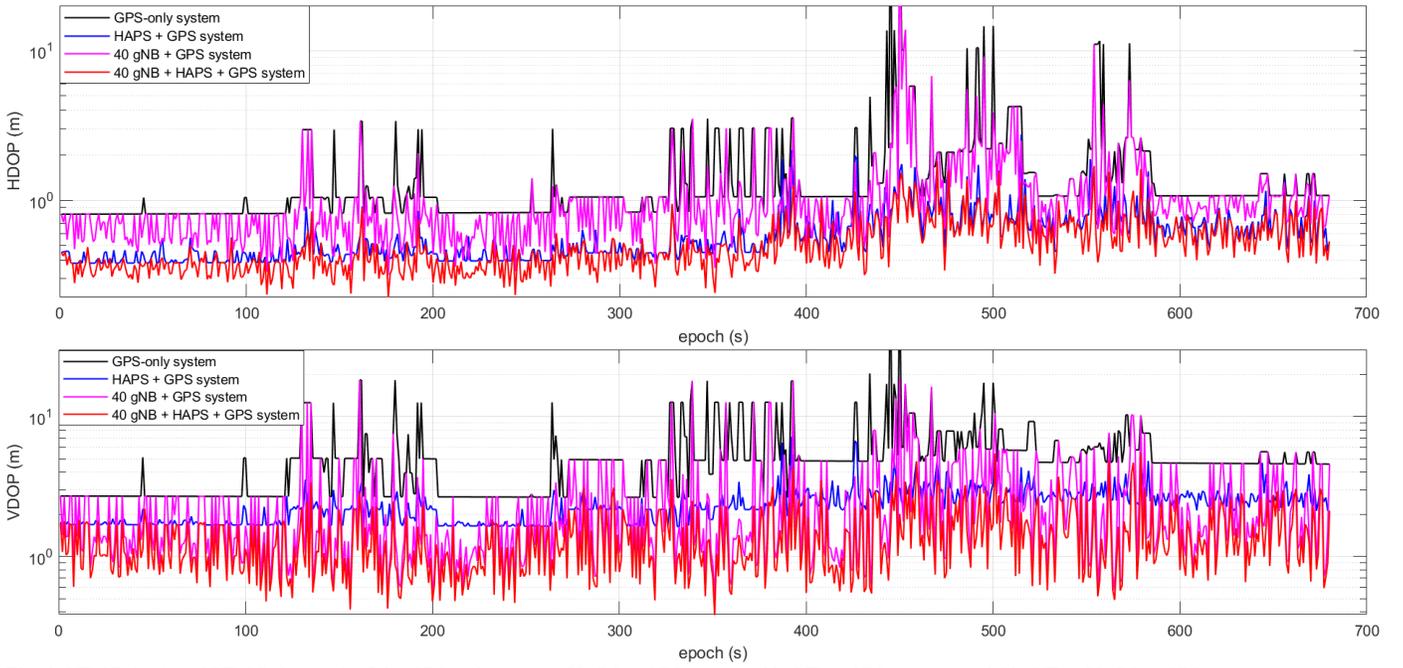

Fig. 8. HDOP (top) and VDOP (bottom) of the GPS-only system, HAPS + GPS system, 40 gNB + GPS system, and 40 gNB + HAPS + GPS system.

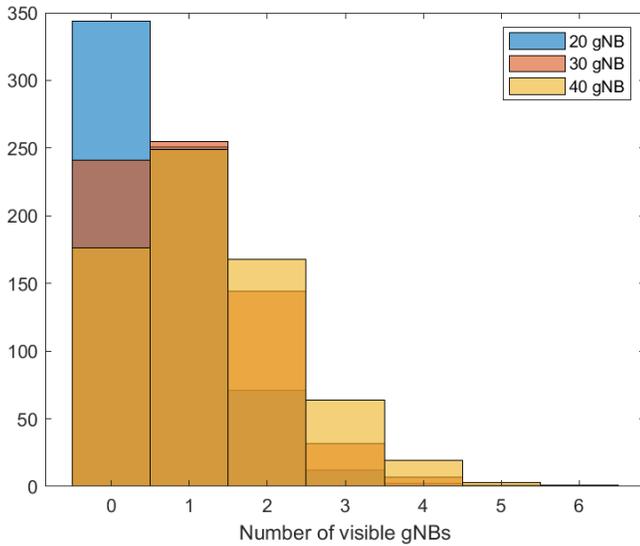

Fig. 7. Histogram of gNB availability for different scenarios with varied density.

TABLE I
STATISTICS ABOUT THE AVAILABILITY OF GNBS BY SCENARIOS

| Number of gNBs | Mean | Median |
| --- | --- | --- |
| 20 | 0.6221 | 0 |
| 30 | 0.9397 | 1 |
| 40 | 1.2456 | 1 |

### III. SIMULATION RESULTS

The availabilities of GPS satellites, HAPS, and gNBs are depicted in Fig. 6. As we can see the availabilities for both HAPS and satellite change abruptly at epoch around 380 s, which is regarded as the boundary in differentiating the suburban and urban areas. As the LOS model for gNB is only concerned with the baseline distance, we do not see an abrupt change in the availability of gNB. Due to the fact that the gNB availabilities for different scenarios with varied density do not differ from each other significantly, it is difficult to capture the availability difference for gNBs in the same epoch-by-epoch availability plot. Therefore, only the 40-gNB scenario along with HAPS and GPS satellite is presented in Fig. 6. A histogram showing the gNB availability for different scenarios with varied density is provided and shown in Fig. 7. Moreover, the statistics about the availability of gNB in different scenarios is summarized in Table I. We can see from Fig. 6 in conjunction with Fig. 7 and Table I, for the 40-gNB scenario, even though the number of available gNBs can reach four and even six occasionally, the average number is still below two. The availability of gNBs is even worse for the 20-gNB and 30-gNB scenarios. The two epochs circled in the figure illustrate that the number of satellites at these two epochs is insufficient to perform the precise 3D localization, meaning that the GPS-only system fails to provide localization at these two epochs. The HDOP and VDOP are considered as indicators for the horizontal positioning performance and vertical positioning performance, respectively.

The HDOP the VDOP during the entire observation period for different systems considered are presented in Fig. 8. To ensure there is on average at least one available gNB in the positioning system, 40 gNBs are used in conjunction with other vertical components. From Fig. 8, we see that the HDOP of the gNB + HAPS + GPS system outperforms the other systems. One the one hand, we can see the HDOP of the HAPS + GPS system is comparable with that of the gNB + HAPS + GPS system. One the other hand, the HDOP of the gNB + GPS system is in between the GPS-only system and the HAPS + GPS system. This demonstrates that gNBs can improve the HDOP, yet slightly. Nevertheless, gNBs can improve the VDOP significantly. As we can see from Fig. 8 in conjunction with Fig. 6, when there is at least one available gNB, the VDOP of the gNB + GPS system is comparable with the gNB + HAPS + GPS system. We can observe that HAPS can improve both HDOP and VDOP of legacy GNSSs, such as GPS, quite significantly. By integrating HAPS and gNB with satellites, we believe that both the HDOP and VDOP can be significantly improved, leading to a better horizontal and vertical positioning performance.

*A. Impact of the Density of gNBs on the Positioning Performance of the gNB Augmented Positioning Systems*

The cumulative distribution functions (CDFs) of the horizontal and vertical positioning accuracy for both suburban and urban areas are presented in Fig. 9 to Fig. 12. We can observe that both the horizontal and vertical positioning accuracies are improved with the increasing number of gNBs. We should note that the average numbers of the available gNBs for different densities of gNBs do not distinct from one another significantly, thence we can expect to see a more significant improvement in positioning performance if the average number of available gNBs is more than one. Since most of the gNB signals are received from low elevation angles, we can expect gNBs to be more effective in improving the vertical positioning accuracy over the horizontal positioning accuracy. As we can see, in suburban areas, although gNBs are able to improve the horizontal positioning accuracy for both the GPS-only system and the HAPS-aided GPS system, the improvement at 95-percentile is negiligle regardless of the number of gNBs. Moreover, in urban areas, the horizontal positioning performance of the gNB + GPS system is comparable with that of the GPS-only system, and the horizontal positioning performance of the HAPS + GPS system is comparable with that of the gNB + HAPS + GPS system regardless of the number of gNBs. However, we can see from Fig. 11 and Fig. 12 that the vertical positioning performance of the gNB + GPS system is much better than the GPS-only system and is comparable with the HAPS + GPS system in both the suburban and urban areas. This demonstrates the effectiveness of gNBs in improving the vertical positioning performance. Note that the number of available gNBs is fewer than the number of available HAPS during the entire observation period, we can expect the vertical positioning performance of the gNB + GPS system to be better than that of the HAPS+ GPS system, if there were a few more available gNBs. On the other hand, we can see that HAPS is able to improve both the horizontal and vertical positioning performance of legacy GNSSs significantly in both suburban and urban areas.

TABLE II
EFFECTIVENESS OF GNBS ON THE GNB AUGMENTED POSITIONING SYSTEMS

| Type of region | | GPS-only system | HAPS + GPS system |
|---|---|---|---|
| Horizontal | Suburban | 10% | 5% |
| | Urban | 0% | 12% |
| Vertical | Suburban | 47% | 27% |
| | Urban | 52% | 52% |

*B. Impact of the Pseudorange Error of gNB on the Positioning Performance of the gNB Augmented Positioning Systems*

To better analyze the impact of the pseudorange error of gNB on positioning performance of the gNB augmented positioning systems, in this section, we only consider the 40-gNB scenario where there is at least one available gNB on average for positioning. Using three different pseudorange errors for gNB, the CDFs of the horizontal positioning accuracy for both suburban and urban areas are presented in Fig. 13 and Fig. 14, respectively; and the CDFs of the vertical positioning accuracy for both suburban and urban areas are presented in Fig. 15 and Fig. 16, respectively. From these figures, we observe that the impact of distinct pseudorange errors of gNB on both horizontal and vertical positioning performance is negligible. This is because the number of available gNBs is few compared to HAPS and satellites. However, we can observe that gNBs are more effective in improving the vertical positioning accuracy than the horizontal positioning accuracy. As we can see, for the 90-percentile horizontal positioning accuracy, with the assistance of gNBs, the GPS-only system is optimized by about

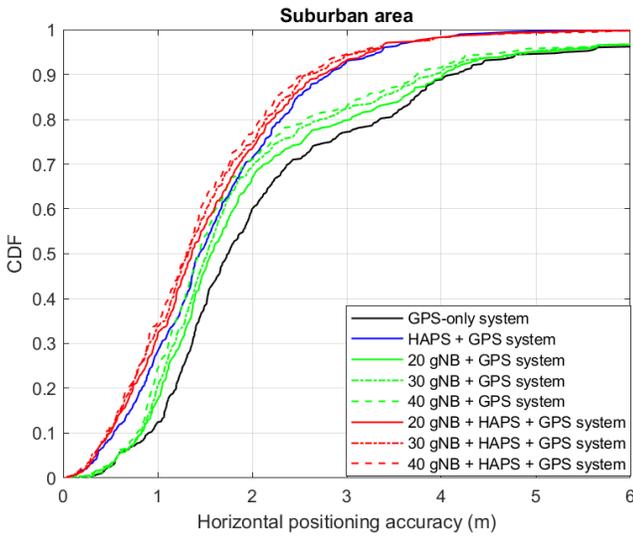

Fig. 9. CDF of the horizontal positioning accuracy for different number of gNB (suburban).

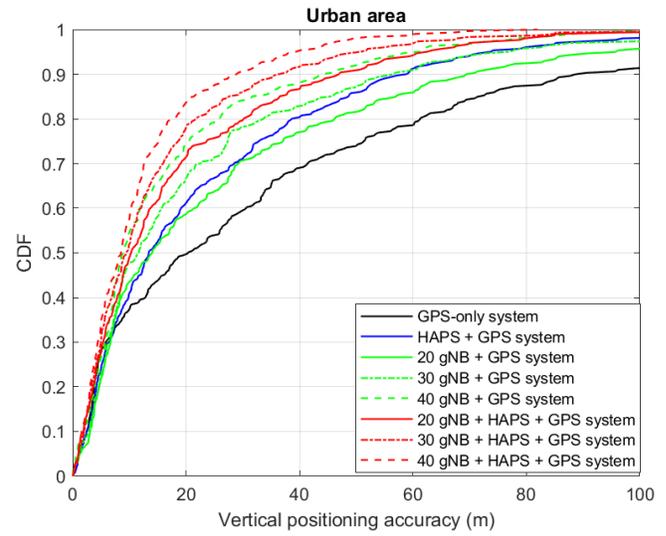

Fig. 12. CDF of the vertical positioning accuracy for different number of gNB (urban).

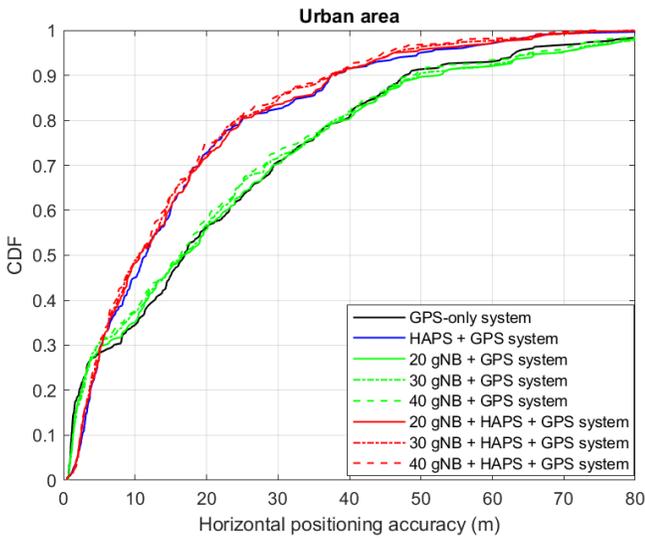

Fig. 10. CDF of the horizontal positioning accuracy for different number of gNB (urban).

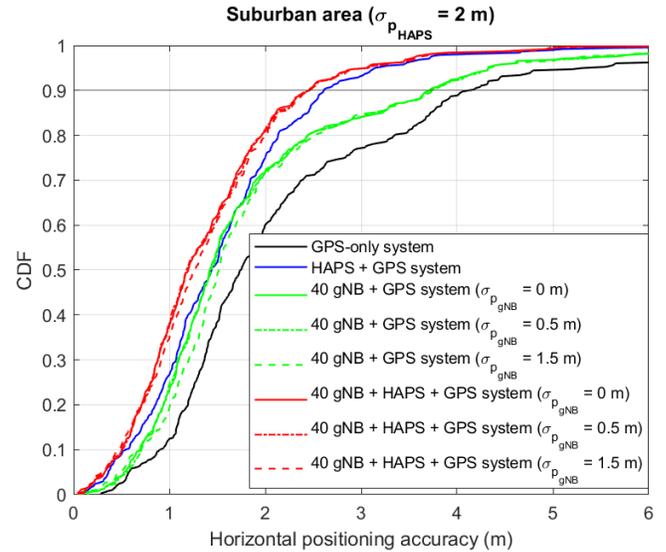

Fig. 13. CDF of the horizontal positioning accuracy for different pseudorange errors of gNB (suburban).

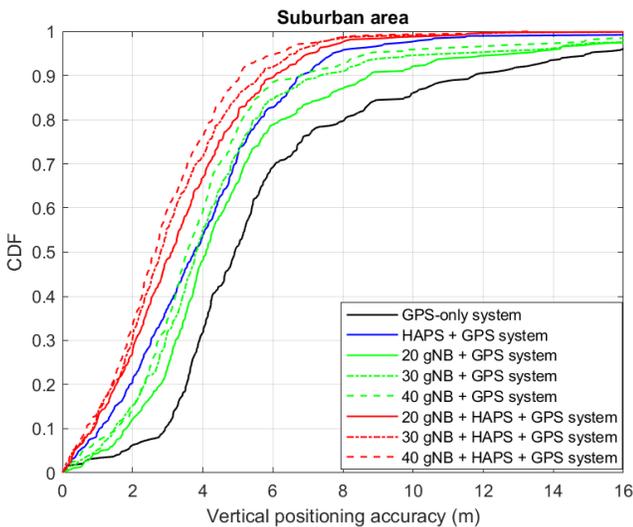

Fig. 11. CDF of the vertical positioning accuracy for different number of gNB (suburban).

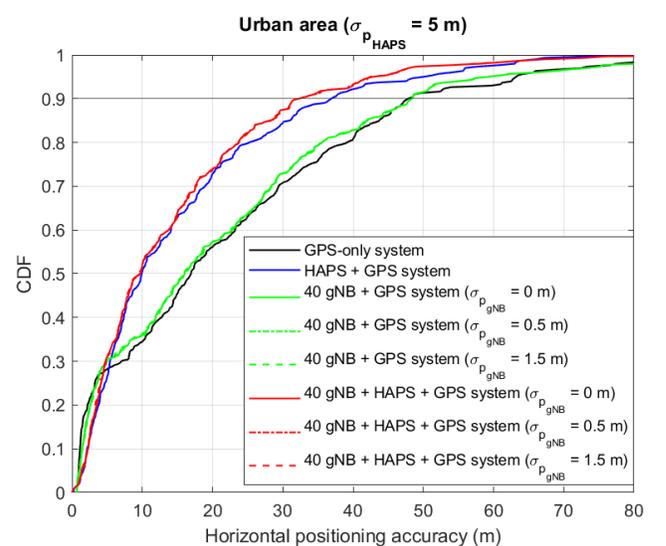

Fig. 14. CDF of the horizontal positioning accuracy for different pseudorange errors of gNB (urban).

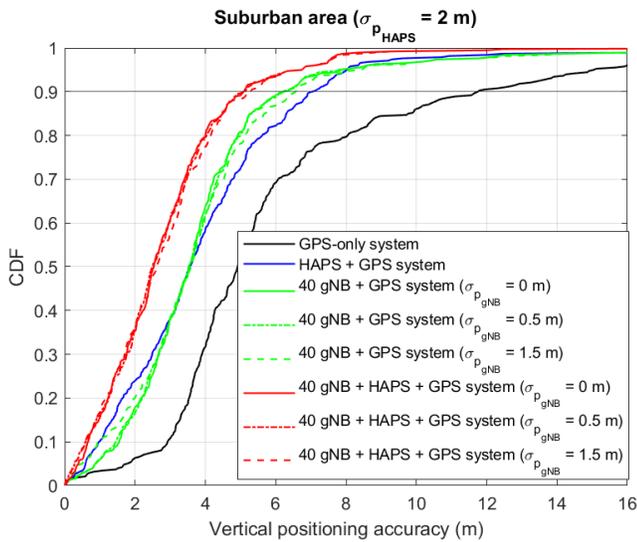

Fig. 15. CDF of the vertical positioning accuracy for different pseudorange errors of gNB (suburban).

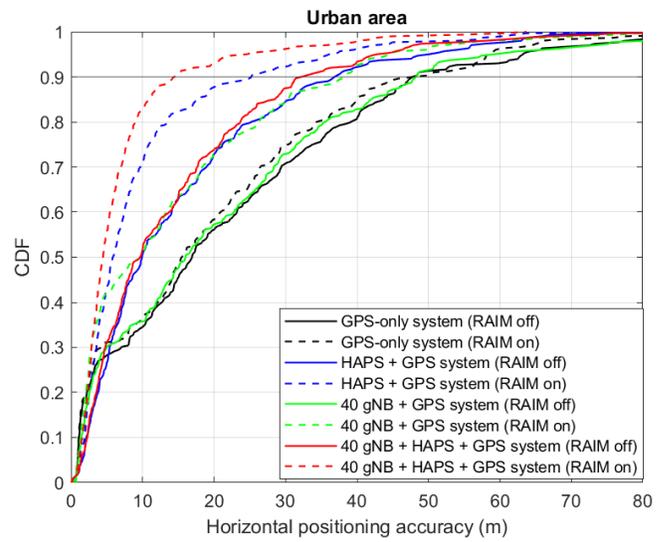

Fig. 18. CDF of the horizontal positioning accuracy with RAIM applied (urban).

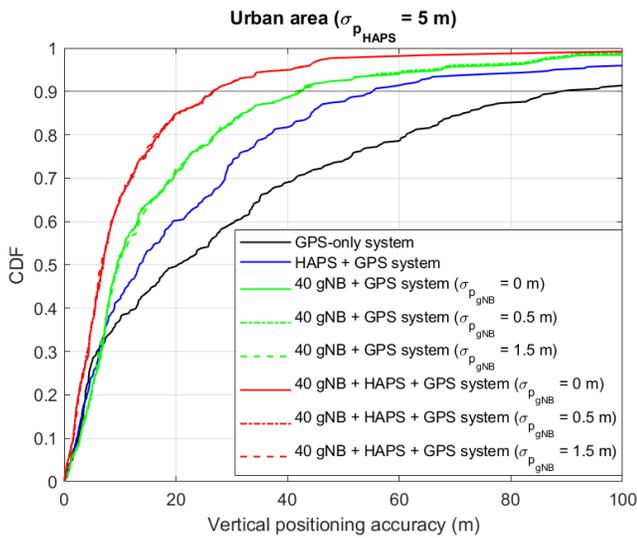

Fig. 16. CDF of the vertical positioning accuracy for different pseudorange errors of gNB (urban).

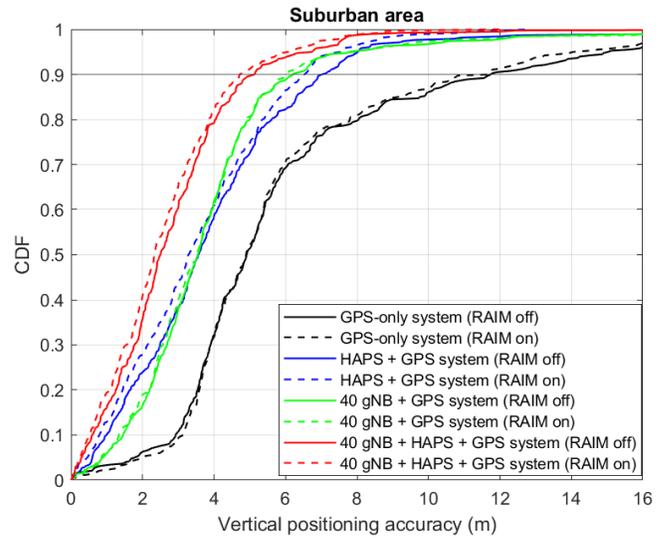

Fig. 19. CDF of the vertical positioning accuracy with RAIM applied (suburban).

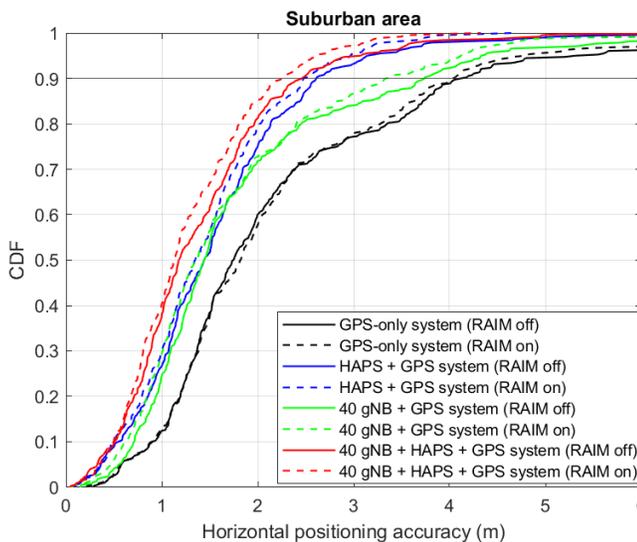

Fig. 17. CDF of the horizontal positioning accuracy with RAIM applied (suburban).

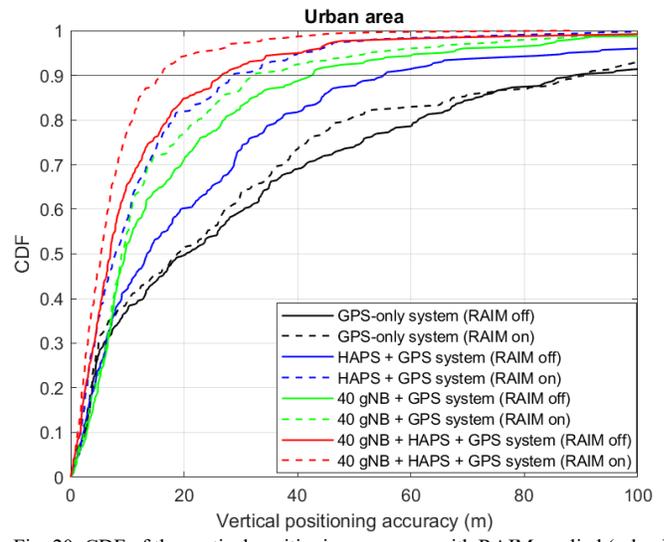

Fig. 20. CDF of the vertical positioning accuracy with RAIM applied (urban).

TABLE III
EFFECTIVENESS OF RAIM ON THE HAPS AND/OR GNB AIDED GPS SYSTEMS

|  | Type of metric | gNB + GPS system | HAPS + GPS system | gNB + HAPS + GPS system |
|---|---|---|---|---|
| **Suburban** | horizontal | 9.79% | 4.35% | 8.2% |
|  | vertical | 2.68% | 7.60% | 7.53% |
| **Urban** | horizontal | 20.98% | 31.58% | 55.17% |
|  | vertical | 24.55% | 48.27% | 40.48% |

10% and 0% in the suburban and urban areas, respectively; the HAPS + GPS system is optimized by about 5% and 12% in the suburban and urban areas, respectively. On the other hand, for the 90-percentile vertical positioning accuracy, the GPS-only system is optimized by about 47% and 52% in the suburban and urban areas, respectively; the HAPS + GPS system is optimized by about 27% and 52% in the suburban and urban areas, respectively. A summary of the effectiveness of gNBs on the gNB augmented positioning systems is provided in Table II.

## C. Effectiveness of RAIM on the HAPS and/or gNB aided GPS Systems

As the region of interest is urban areas where signal quality can be severely degraded, we would like to investigate the effectiveness of RAIM algorithms on three hybrid systems: the HAPS + GPS system, the gNB + GPS system, and the gNB + HAPS + GPS system. As the difference among the curves of the same system with distinct gNB pseudorange errors is negligible, we choose to use the medium value of the pseudorange error for gNB in this section. Fig. 17 and Fig. 18 show the CDFs of the horizontal positioning accuracy for both suburban and the urban areas, respectively; and Fig. 19 and Fig. 20 show the CDFs of the vertical positioning accuracy in both suburban and urban areas, respectively. As we can see, in suburban areas, the 90-percentile horizontal positioning accuracy of the gNB + GPS system is optimized by about 9.79%; and the 90-percentile vertical positioning accuracy of the gNB + GPS system is optimized by about 2.68%, with RAIM enabled. The 90-percentile horizontal positioning accuracy of the HAPS + GPS system is optimized by about 4.35%; and the 90-percentile vertical positioning accuracy of the HAPS + GPS system is optimized by about 7.60%, with RAIM enabled. For the gNB + HAPS + GPS system, the 90-percentile horizontal positioning accuracy is optimized by about 8.20%; and the 90-percentile vertical positioning accuracy is optimized by about 7.53%, with RAIM enabled. By contrast, RAIM is much more effective in improving the positioning performance in the urban area. On the other hand, in urban areas, we see that the 90-percentile horizontal positioning accuracy of the gNB + GPS system is optimized by about 20.98%; and the 90-percentile vertical positioning accuracy of the gNB + GPS system is optimized by about 24.55%, with RAIM enabled. The 90-percentile horizontal positioning accuracy of the HAPS + GPS system is optimized by about 31.85%; and the 90-percentile vertical positioning accuracy of the HAPS + GPS system is optimized by about 48.27%, with RAIM enabled. For the gNB + HAPS + GPS system, the 90-percentile horizontal positioning accuracy is optimized by about 55.17%; and the 90-percentile vertical positioning accuracy is optimized by about 40.48%, with RAIM enabled. A summary of the effectiveness of RAIM on the HAPS and/or gNB aided GPS systems is provided in Table III. As we can see, RAIM is not very effective in improving the positioning performance in the suburban area where the signals are usually of good quality. Due to the few available satellites in the urban area, we observe that RAIM is not very effective on the GPS-only system. For a similar reason that the number of available gNBs is few, we observe that RAIM is less effective on the gNB + GPS system compared to the HAPS + GPS system and the gNB + HAPS + GPS system. This shows that the effectiveness of RAIM is subject to the number of ranging sources.

## IV. CONCLUSION

Undoubtedly, satellites are essential in providing broadband communication services, such as weather broadcast, geodesy, and localization due to the superior coverage. However, they are inadequate in providing highly accurate and reliable positioning services owing to high pathloss, high latency, and prominent atmospheric effect. In the urban areas, the number of available satellites might be insufficient to carry out a precise 3D localization. One way to enhance the positioning performance of legacy GNSSs is to deploy HAPS as additional ranging sources that are equipped with satellite-grade atomic clocks. Thanks to many advantages offered by the HAPS, such as smaller time dilation, shorter distance to the ground, and stationarity, the quality of HAPS signals will likely be better than that of satellite signals. Apart from HAPS, we can utilize gNB signals to improve the positioning performance, especially in the vertical direction. Thanks to the proximity and the low elevation angle of terrestrial signals, gNBs are more effective in improving the VDOP over the HDOP. In this paper, we demonstrated the effectiveness of gNBs in improving the vertical positioning accuracy and HAPS in improving both the horizontal and vertical positioning accuracy. We also demonstrated the effectiveness of RAIM on the HAPS and/or gNB aided GPS systems in urban areas.

To further enhance the urban outdoor positioning performance, we can equip high-definition (HD) cameras under HAPS. Since HAPS are positioned at a lower altitude compared to satellites, the images sent from HAPS will exhibit a higher resolution compared to that sent from satellites. Images, which are considered as soft contextual information (SCI), can be utilized to enhance the positioning performance. Optimized signal design, such as GPS L5, can be an effective solution in mitigating the multipath effect, thereby further optimizing the positioning performance in urban areas.

## ACKNOWLEDGMENT

We would like to express our gratitude to Orolia for the Skydel GNSS software donated to Carleton University.

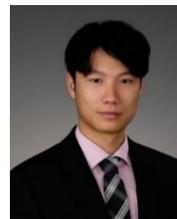

**Hongzhao Zheng** (Member, IEEE) received the B. Eng. (Hons.) degree in engineering physics from the Carleton University, Ottawa, ON, Canada, in 2019. He is currently a PhD student at Carleton University. His research interest is urban positioning using sensor-enabled heterogeneous wireless infrastructure.

Mr. Zheng received the best paper award at IEEE WiSEE 2022.

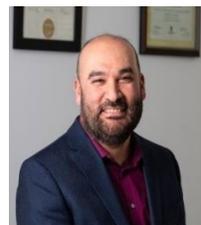

**Mohamed Atia** (Senior Member, IEEE) received the B.S. and M.Sc. degrees in computer systems from Ain Shams University, Cairo, Egypt, in 2000 and 2006, respectively, and the Ph.D. degree in electrical and computer engineering from Queen's University, Kingston, ON, Canada, in 2013. He is currently an Associate Professor with the Department of Systems and Computer Engineering, Carleton University. He is also the Founder and the Director of the Embedded and Multi-Sensory Systems Laboratory (EMSLab), Carleton University. His research interests include sensor fusion, navigation systems, artificial intelligence, and robotics.

Dr. Atia received the best paper award at IEEE WiSEE 2022.

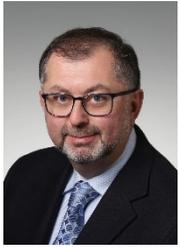

**Halim Yanikomeroglu** (Fellow, IEEE) received the BSc degree in electrical and electronics engineering from the Middle East Technical University, Ankara, Turkey, in 1990, and the MASc degree in electrical engineering (now ECE) and the PhD degree in electrical and computer engineering from the University of Toronto, Canada, in 1992 and 1998, respectively. Since 1998 he has been with the Department of Systems and Computer Engineering at Carleton University, Ottawa, Canada, where he is now a Full Professor.

Dr. Yanikomeroglu's research interests cover many aspects of wireless communications and networks, with a special emphasis on non-terrestrial networks (NTN) in the recent years. He has given 110+ invited seminars, keynotes, panel talks, and tutorials in the last five years. He has supervised or hosted over 150 postgraduate researchers in his lab at Carleton. Dr. Yanikomeroglu's extensive collaborative research with industry resulted in 39 granted patents. Dr. Yanikomeroglu is a Fellow of the IEEE, the Engineering Institute of Canada (EIC), and the Canadian Academy of Engineering (CAE). He is a Distinguished Speaker for the IEEE Communications Society and the IEEE Vehicular Technology Society, and an Expert Panelist of the Council of Canadian Academies (CCA|CAC).

Dr. Yanikomeroglu is currently serving as the Chair of the Steering Committee of IEEE's flagship wireless event, Wireless Communications and Networking Conference (WCNC). He is also a member of the IEEE ComSoc Governance Council, IEEE ComSoc GIMS, IEEE ComSoc Conference Council, and IEEE PIMRC Steering Committee. He served as the General Chair and Technical Program Chair of several IEEE conferences. He has also served in the editorial boards of various IEEE periodicals.

Dr. Yanikomeroglu received several awards for his research, teaching, and service, including the IEEE ComSoc Fred W. Ellersick Prize (2021), IEEE VTS Stuart Meyer Memorial Award (2020), and IEEE ComSoc Wireless Communications TC Recognition Award (2018). He received best paper awards at IEEE Competition on Non-Terrestrial Networks for B5G and 6G in 2022 (grand prize), IEEE ICC 2021, IEEE WISEE 2021 and 2022.